\documentclass[aps,prb,twocolumn,superscriptaddress,amsmath,amssymb]{revtex4-2}
\usepackage{amssymb}
\usepackage{graphicx}
\usepackage{tabularx}
\usepackage{bm}
\usepackage{epsfig}
\usepackage{graphicx}
\usepackage{rotating}
\usepackage[ansinew]{inputenc}

\begin{document}

\title{Tuning the spontaneous exchange bias effect in La$_{1.5}$Sr$_{0.5}$CoMnO$_6$ with sintering temperature}

\author{C. Macchiutti}
\affiliation{Centro Brasileiro de Pesquisas F\'{\i}sicas, Rio de Janeiro, RJ 22290-180, Brazil}
\author{J. R. Jesus}
\affiliation{Centro Brasileiro de Pesquisas F\'{\i}sicas, Rio de Janeiro, RJ 22290-180, Brazil}
\affiliation{Universidade Federal do Rio de Janeiro, Rio de Janeiro, RJ 21941-853, Brazil}
\author{F. B. Carneiro}
\affiliation{Centro Brasileiro de Pesquisas F\'{\i}sicas, Rio de Janeiro, RJ 22290-180, Brazil}
\author{L. Bufai\c{c}al}
\email{lbufaical@ufg.br}
\affiliation{Instituto de F\'{\i}sica, Universidade Federal de Goi\'{a}s, Goi\^{a}nia, GO 74001-970, Brazil}
\author{R. A. Klein}
\affiliation{Materials, Chemical, and Computational Sciences, National Renewable Energy Laboratory, Golden, CO 80401, USA}
\affiliation{NIST Center for Neutron Research, National Institute of Standards and Technology, Gaithersburg, MD 20899, USA}
\author{Q. Zhang}
\affiliation{Neutron Scattering Division, Oak Ridge National Laboratory, Oak Ridge, TN 37831, USA}
\author{M. Kirkham}
\affiliation{Neutron Scattering Division, Oak Ridge National Laboratory, Oak Ridge, TN 37831, USA}
\author{C. M. Brown}
\affiliation{NIST Center for Neutron Research, National Institute of Standards and Technology, Gaithersburg, MD 20899, USA}
\affiliation{Department of Chemical Engineering, University of Delaware; Newark, DE 19716, USA}
\author{R. D. dos Reis}
\affiliation{Laborat\'{o}rio Nacional de Luz S\'{\i}ncrotron, Centro Nacional de Pesquisa em Energia e Materiais, Campinas, SP 13083-970, Brazil}
\author{G. Perez}
\affiliation{Departamento de Engenharia Mec\^{a}nica, Universidade Federal Fluminense, Niter\'{o}i, RJ 24210-240, Brazil}
\author{E. M. Bittar}
\email{bittar@cbpf.br}
\affiliation{Centro Brasileiro de Pesquisas F\'{\i}sicas, Rio de Janeiro, RJ 22290-180, Brazil}

\begin{abstract}
Here, we present a study of the influence of microstructure on the magnetic properties of polycrystalline samples of the La$_{1.5}$Sr$_{0.5}$CoMnO$_6$ double perovskite, with primary attention to the spontaneous exchange bias effect, a fascinating recently discovered phenomena for which some materials exhibit unidirectional magnetic anisotropy after being cooled in zero magnetic fields. By sintering La$_{1.5}$Sr$_{0.5}$CoMnO$_6$ at different temperatures, we obtained samples with distinct average grain sizes, ranging from 1.54 to 6.65 $\mu$m. A detailed investigation of the material's structural, morphologic, electronic, and magnetic properties using X-ray powder diffraction, powder neutron diffraction, X-ray absorption near edge structure spectroscopy, scanning electron microscopy, and AC and DC magnetometry has revealed a systematic enhancement of the exchange bias effect with increasing the average grain size. This evolution is discussed in terms of changes in the material's porosity and grain morphology and its influence on the exchange couplings at the magnetic interfaces.
\end{abstract}

\maketitle

\section{Introduction}
Transition-metal-based oxides usually exhibit strong electronic correlation and complex interplay between charge, spin, lattice, and orbital degrees of freedom that can lead to a plethora of interesting physical properties such as multiferroicity, high-temperature superconductivity, metal-insulator transitions, giant magnetoresistance, and many others \cite{Rao,Tokura,Sami}. A fascinating effect observed for some (but not restricted to) polycrystalline oxides is the exchange bias (EB) effect, a phenomenon of unidirectional magnetic anisotropy (UA) set by uncompensated exchange interactions at the interfaces of distinct magnetic phases present in the material, manifested by a shift in the curve of magnetization as a function of applied magnetic field [$M(H)$] \cite{Nogues}. It was firstly observed in polycrystalline nanoparticles consisting of a ferromagnetic (FM) core of Co embedded on a CoO antiferromagnetic (AFM) shell \cite{Meiklejohn}, but later on it was also found on FM-ferrimagnetic, AFM-ferrimagnetic, FM-spin glass (SG) and other heterogeneous systems \cite{Nogues,Nogues2}.

The EB effect finds its applicability in spin valves, magnetic recording heads, and tunneling devices; thus, the underlying mechanisms responsible for its exchange anisotropy have been widely debated over the last decades. In the case of polycrystalline oxides, the role of grain size on the EB was systematically studied. The results show that there is not a simple direct relation between grain size and the magnitude of the EB effect, resulting from the fact that changes in grain size affect several parameters such as the spin structure, the texture, the anisotropy, and others, and the influence of each parameter on the EB may vary from one material to another \cite{Nogues,Nogues2}.

Usually, the UA of an EB material emerges after cooling it in the presence of an external magnetic field ($H$). However, recently some compounds were found for which the UA is set spontaneously, \textit{i.e.} it occurs even without the assistance of a cooling field \cite{Wang,Maity}. This is the so-called spontaneous exchange bias (SEB) effect, which is characterized by the fact that all the SEB materials discovered so far have a glassy magnetic state in common at low temperatures. It has already been shown that the unusually long-lasting relaxation of the SG-like moments plays a vital role in the asymmetry of the $M(H)$ curves of SEB systems \cite{Model,Model2}. In this context, the double-perovskite oxides with general formula A$_{2}$BB'O$_6$ (A = rare-earth/alkaline-earth, B and B' = transition-metal ions) stand as prospective candidates to exhibit SEB since the cationic disorder at the B/B' sites and the competing magnetic interactions caused by the presence of distinct transition-metal ions often lead to SG-like behavior \cite{Sami,Rev_SG}.

The first observation of SEB on a double-perovskite oxide was on the La$_2$CoMnO$_6$ polycrystal doped with 25\% of Sr at the La site \cite{Murthy}. Regarding the magnetic interactions on  La$_{1.5}$Sr$_{0.5}$CoMnO$_6$ (LSCMO), besides the Co$^{2+}$--O--Mn$^{4+}$ FM coupling typically found in the La$_2$CoMnO$_6$ parent compound, the partial substitution of La$^{3+}$ by Sr$^{2+}$ leads to mixed valence states on the transition-metal ions, resulting in additional exchange interactions that, together with the antisite disorder at the Co/Mn site, drive the system to a SG behavior at temperatures below the ordinary FM/AFM transitions, making this a reentrant spin glass material that shows a substantial SEB effect \cite{Murthy,Murthy2}.

After the observation of SEB on LSCMO, other similar polycrystalline double-perovskite oxides such as La$_{2-x}$A$_x$CoMnO$_6$ with A = Ba, Ca, Sr, La$_{1.5}$Sr$_{0.5}$Co$_{1-x}$B$_x$MnO$_6$ with B = Fe, Ga, Pr$_{2-x}$Sr$_x$CoMnO$_6$, Sm$_{1.5}$Ca$_{0.5}$CoMnO$_6$ were also discovered to exhibit spontaneous UA \cite{La2-xCax,PRB2019,APL,La2-xBax,Zhang,JPCM,Zhao,Pal,Giri}. At the same time, it was recently demonstrated that single crystalline samples of LSCMO and La$_{1.5}$Ca$_{0.5}$CoMnO$_6$ do not show SG-like behavior nor SEB effect \cite{single_crystal}, further evidencing that glassy magnetism is required for the appearance of SEB on double perovskites. Nevertheless, it remains a puzzle that competing magnetic phases leading to an SG-like state are shared among several double-perovskite oxides. Still, most of them, including CoMn-based ones, do not show any trace of SEB effect \cite{Sami}. So, besides the SG-like behavior, what are the other fundamental parameters responsible for the emergence of spontaneous UA? At first glance, one may suspect the material's morphology since a striking difference between single and polycrystals is that in the former, the order extends, in principle, over the entire volume, although in practice, it may be divided into domains. Simultaneously, for polycrystals, the order only exists over small regions of the crystal, the grains, which can be further divided into even smaller domains whose boundaries usually present defects, frustration, spin canting, etc.

The previous characterization of LSCMO makes this material the ideal playground for further study on the underlying physics responsible for the SEB. In this work, we use this compound as a prototype to investigate the influence of microstructure on the spontaneous UA observed on double-perovskite oxides. We synthesized polycrystalline LSCMO at different sintering temperatures to produce samples with distinct average grain sizes and coalescence. By using powder X-ray diffraction (PXRD), powder neutron diffraction (PND), X-ray absorption near edge structure (XANES) spectroscopy, scanning electron microscopy (SEM), and AC and DC magnetization techniques, we were able to get a deeper insight into the mechanisms involved in the evolution of the UA and its correlation with grain size and boundaries.

\section{Experimental details}
The four polycrystalline samples of LSCMO here investigated were produced by conventional solid-state reaction at four distinct sintering temperatures: $1100^{\circ}$C, $1200^{\circ}$C, $1300^{\circ}$C, and $1400^{\circ}$C. Each sample will be hereafter called by its sintering temperature. The synthesized samples underwent a thermal treatment methodology described as follows. Initially, precursors of La$_2$O$_3$, MnO, Co$_3$O$_4$, and SrCO$_3$ with a purity higher than 99.9\% (Sigma Aldrich) were used as starting materials, weighed in stoichiometric ratios and then mixed in a mortar until complete homogenization. Subsequently, a thermal treatment routine was established, where the sample was heated/cooled to/from 1100$^{\circ}$C at a rate of 3$^{\circ}$C/min after 48 hours at the maximum temperature. This procedure was repeated eight times, always performing homogenization in an agate mortar and sieving of the powder between each turn. After obtaining a homogenous powder in the desired crystallographic structure, the pellets were sintered for 24 hours at four different temperatures (1100$^{\circ}$C, 1200$^{\circ}$C, 1300$^{\circ}$C, and 1400$^{\circ}$C), always heated/cooled to/from the maximum temperature at a rate of 3$^{\circ}$C/min.

The structural properties of the samples were investigated by PXRD and PND. PXRD data were measured using a PANalytical Empyrean diffractometer with Cu-$K_\alpha$ radiation ($\lambda=1.5406$ \AA). All data were collected in Bragg-Brentano geometry in the continuous mode with a 2$\theta$ range from 20$^\circ$ to 80$^\circ$, with a step size of 0.013$^\circ$, and a scanning speed of 0.5$^\circ$/min. PND data were collected using frame 1 ($\lambda$ centered around 0.8 \AA, $d$ coverage from 0.1-8 \AA) at the POWGEN time-of-flight powder diffractometer at the Spallation Neutron Source, Oak Ridge National Laboratory. One pattern was collected for sample 1300 using frame 2 ($\lambda$ centered around 1.5 \AA). We analyzed the data using the TOPAS Academic software package \cite{TOPAS} in conjunction with the CMPR software package \cite{CMPR}. Le Bail/Pawley fits using these data were initialized using the lattice parameters derived from the fits of the PXRD pattern for each sample.

The morphological analysis of the samples was performed with a field emission gun Jeol JEM 7100FT high-resolution scanning electron microscope (SEM) operated at 15 kV in secondary electrons mode. The grain size of the images was obtained using the ImageJ software. To improve statistics, different axes of the particles were considered.

XANES spectroscopy experiments at ambient conditions at Co and Mn $K$-edges were performed in transmission geometry at the Extreme Methods of Analysis (EMA) beamline of the 4th generation Brazilian synchrotron, Sirius \cite{LNLS}. Further details of the XANES experiments can be found in the Supplementary Material at Ref. \onlinecite{SM}.

Magnetic data were collected on a Quantum Design PPMS-VSM magnetometer. DC magnetization against temperature was measured at zero-field-cooled (ZFC) and field-cooled (FC) modes. AC magnetic susceptibility was measured with driving field $H_{ac}=5$ Oe at the frequency range of 100-10 000 Hz. All $M(H)$ loops were performed at $T=5$ K up to a maximum magnetic field of $H_{max}=\pm90$ kOe after a 300-minute wait time to guarantee thermal stabilization.

\section{Results}
\begin{figure}
\begin{center}
\includegraphics[width=0.49 \textwidth]{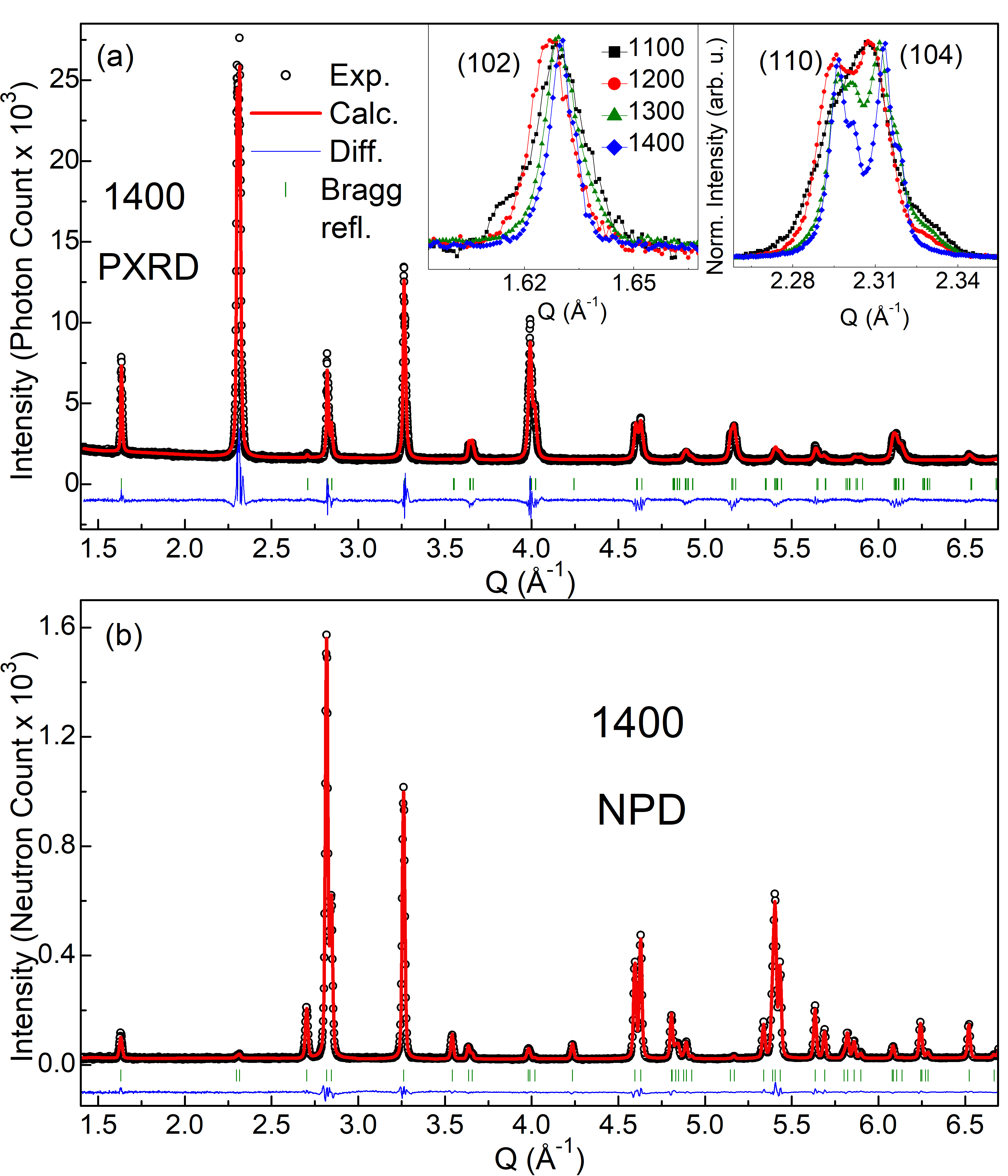}
\end{center}
\caption{Rietveld refinement fittings of (a) PXRD and (b) PND patterns of LSCMO sample 1400. The vertical bars represent the Bragg reflections for the $R\bar{3}c$ space group. The upper insets show magnified views of the PXRD patterns for all samples at the (102), (110), and (104) reflections.}
\label{Fig_XNPD}
\end{figure}

The PXRD patterns attest to the crystallinity and bulk purity of the samples, all belonging to the rhombohedral $R\bar{3}c$ space group, in agreement with previous reports \cite{Murthy,Murthy2}. While the Rietveld refinements resulted in nearly the same lattice parameters for all the samples, magnified views of the (102), (110), and (104) Bragg reflections depicted in the insets of Fig. \ref{Fig_XNPD}(a) reveal that the peaks get sharper and more defined as the sintering temperature ($T_s$) increases, suggesting that higher $T_s$ leads to larger grain size and higher crystallinity.

The similar X-ray scattering cross sections for Co and Mn in conventional Cu-$K_{\alpha}$ radiation preclude an accurate assessment of the antisite disorder in LSCMO with PXRD alone. Since the antisite disorder at the transition-metal ion sites is known to impact the electronic and magnetic properties of perovskites directly \cite{Sami,Serrate}, and given the significantly different coherent neutron scattering lengths for $^{nat}$Mn and $^{nat}$Co (-3.73 fm and 2.49 fm, respectively \cite{Sears}), we conducted room temperature PND measurements on all samples to determine the extent of antisite disorder. Fig. \ref{Fig_XNPD}(b) shows the PND pattern collected using $\lambda$ centered around 0.8 \AA ($d$ coverage from 0.1 to 8 \AA) for sample 1400 as representative of all the others, for which the same overall behavior was found. In the Supplementary Material is shown the data analysis for all samples \cite{SM}.

We do not detect any reflections corresponding to rocksalt double-perovskites in the PND patterns, in which the transition-metal ions would systematically alternate positions. Instead, our results confirm that Co and Mn are disordered at the centers of the oxygen octahedra. For sample 1300, we have additionally measured PND using $\lambda$ centered around 1.5 \AA ($d$ coverage from 0.5 to 12.5 \AA), and no additional peaks were observed in this d-spacing range with respect to the data collected with $\lambda$ centered around 0.8 \AA.

Simultaneous Rietveld refinements of the PXRD and PND data confirm the similarity between the lattice parameters (see Table \ref{T1}). Furthermore, the occupancies of the Co and Mn ions are nearly the same for all samples. To further test possible trends in the antisite disorder with increasing $T_s$, we attempted refinements of the PND patterns using a modified $R\bar{3}$ structure in which Mn and Co ions were constrained to separate sites. This model gave substantially worse fits to the data, indicating that the data are likely best represented by the $R\bar{3}c$ model, a disordered structure in which Co and Mn share the same crystallographic site. Both the antisite disorder and the occupancy of the transition-metal ions at their sites are known to rule the physical properties of perovskites. In our case, the PXRD and PND results indicate that the LSCMO samples investigated here are nearly stoichiometric (especially the three ones produced with higher $T_s$) and that there are no significant changes in the antisite disorder, suggesting that the evolution of the magnetic properties that will be discussed later is not directly related to changes in the lattice parameters nor to the distribution of Co and Mn along the lattice.

\begin{sidewaystable}
\caption{Main results obtained from the PXRD, PND, and SEM data. Occupancy and isotropic displacement parameters for each atom for the four LSCMO samples as determined by simultaneous Rietveld refinements of the PND and PXRD patterns. Values in parentheses indicate $\pm1\sigma$. Additional information on the structures can be found in the CIFs (CSD deposition numbers 2300498-2300501). Displacement parameters of atoms sharing a site were constrained to one single refining value. *Indicates values from the co-refinement of patterns collected with $\lambda$ centered around 0.8 \AA and 1.5 \AA  simultaneously with the PXRD pattern.}
\label{T1}
\begin{tabular}{c|cccc}
\hline \hline
$T_s$ ($^{\circ}$C) & 1100 & 1200 & 1300* & 1400 \\
\hline
$a$ (\AA) & 5.46306(7) & 5.46809(4) & 5.46788(3) & 5.46910(4) \\
$c$ (\AA) & 13.2830(4) & 13.2638(2) & 13.2636(1) & 13.2523(2) \\
$V$ (\AA$^{3}$) & 343.320(9) & 343.455(5) & 343.423(3) & 343.284(4) \\
\hline
Composition
 & La$_{1.477}$Sr$_{0.500}$Co$_{0.967}$Mn$_{0.987}$O$_{6.000}$ & La$_{1.500}$Sr$_{0.498}$Co$_{0.978}$Mn$_{1.020}$O$_{6.000}$ & La$_{1.500}$Sr$_{0.504}$Co$_{0.984}$Mn$_{1.020}$O$_{5.982}$ & La$_{1.488}$Sr$_{0.510}$Co$_{0.996}$Mn$_{1.002}$O$_{5.988}$ \\
Occupation: \\
La & 0.739(5) & 0.749(4) & 0.749(4) & 0.744(3) \\
Sr & 0.250(5) & 0.250(4) & 0.251(4) & 0.256(3) \\
Co & 0.483(2) & 0.490(2) & 0.491(2) & 0.499(1) \\
Mn & 0.493(2) & 0.509(2) & 0.509(2) & 0.501(1) \\
O & 1.000(1) & 1.000(1) & 0.997(1) & 0.998(1) \\
B$_{eq}$ (\AA$^{2}$): \\
La/Sr & 0.450(3) & 0.586(4) & 0.497(4) & 0.469(2) \\
Co/Mn & 0.08(4) & 0.68(3) & 0.89(3) & 0.35(2) \\
O & 0.651(3) & 0.916(3) & 0.749(5) & 0.787(2) \\
\hline
$R_{wp}$ (\%) & 5.239 & 4.824 & 6.739 & 3.449 \\
GOF & 2.70 & 2.74 & 2.94 & 2.04 \\
\hline
$\langle D \rangle$ ($\mu$m) & 1.54(53) & 1.93(84) & 2.89(118) & 6.65(252) \\
\hline \hline
\end{tabular}
\end{sidewaystable}

\begin{figure}
\begin{center}
\includegraphics[width=0.3 \textwidth]{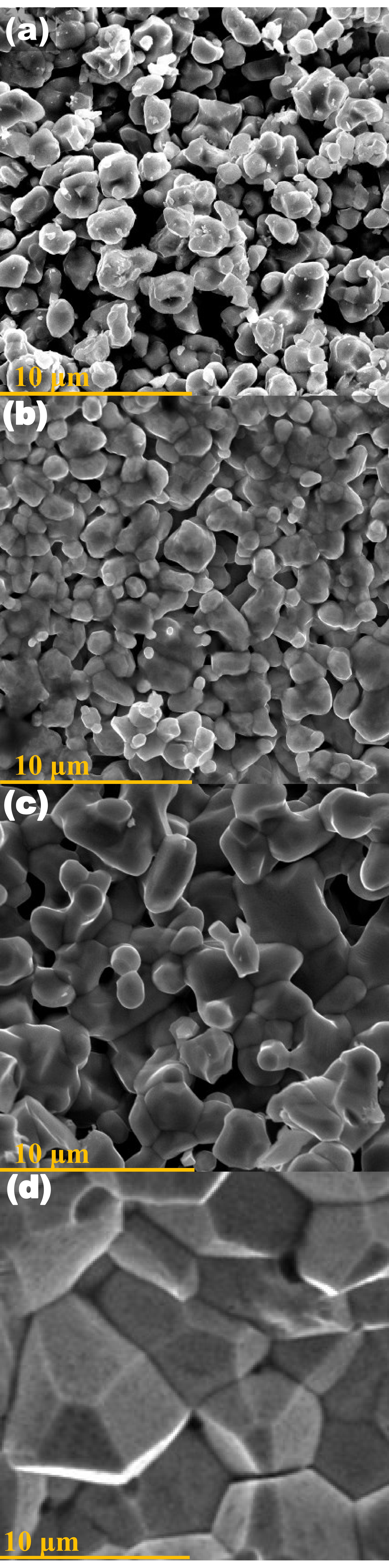}
\end{center}
\caption{SEM images, at 2000 times magnification, 10 mm working distance, and operating at 15 kV, for LSCMO samples (a) 1100, (b) 1200, (c) 1300, and (d) 1400.}
\label{Fig_SEM}
\end{figure}

At the same magnification, the SEM images in Fig. \ref{Fig_SEM} evidence a systematic increase in the grain size with increasing $T_s$ (see Table \ref{T1} and Fig. S12 of the Supplementary Material \cite{SM}). The images also indicate enhancing the system's density, reducing porosity, and exhibiting increased coalescence with $T_s$, typical of the sintering process. The smaller grain size observed for the 1100 sample helps to explain its poorer crystallinity since its grains present a large surface-to-core ratio. The grain boundaries usually exhibit a high degree of frustration, disorder, non-stoichiometry, and other defects \cite{Nogues,Nogues2,Bera}. Changes in the grain's morphology as a function of $T_s$ are also observed in Fig \ref{Fig_SEM}, with the 1400 sample presenting faceted interfaces with impingement of the grains. This further suggests that changes in the exchange coupling at the intergrain interfaces may also occur for the different samples, impacting the material's magnetic properties.

\begin{figure}
\begin{center}
\includegraphics[width=0.5 \textwidth]{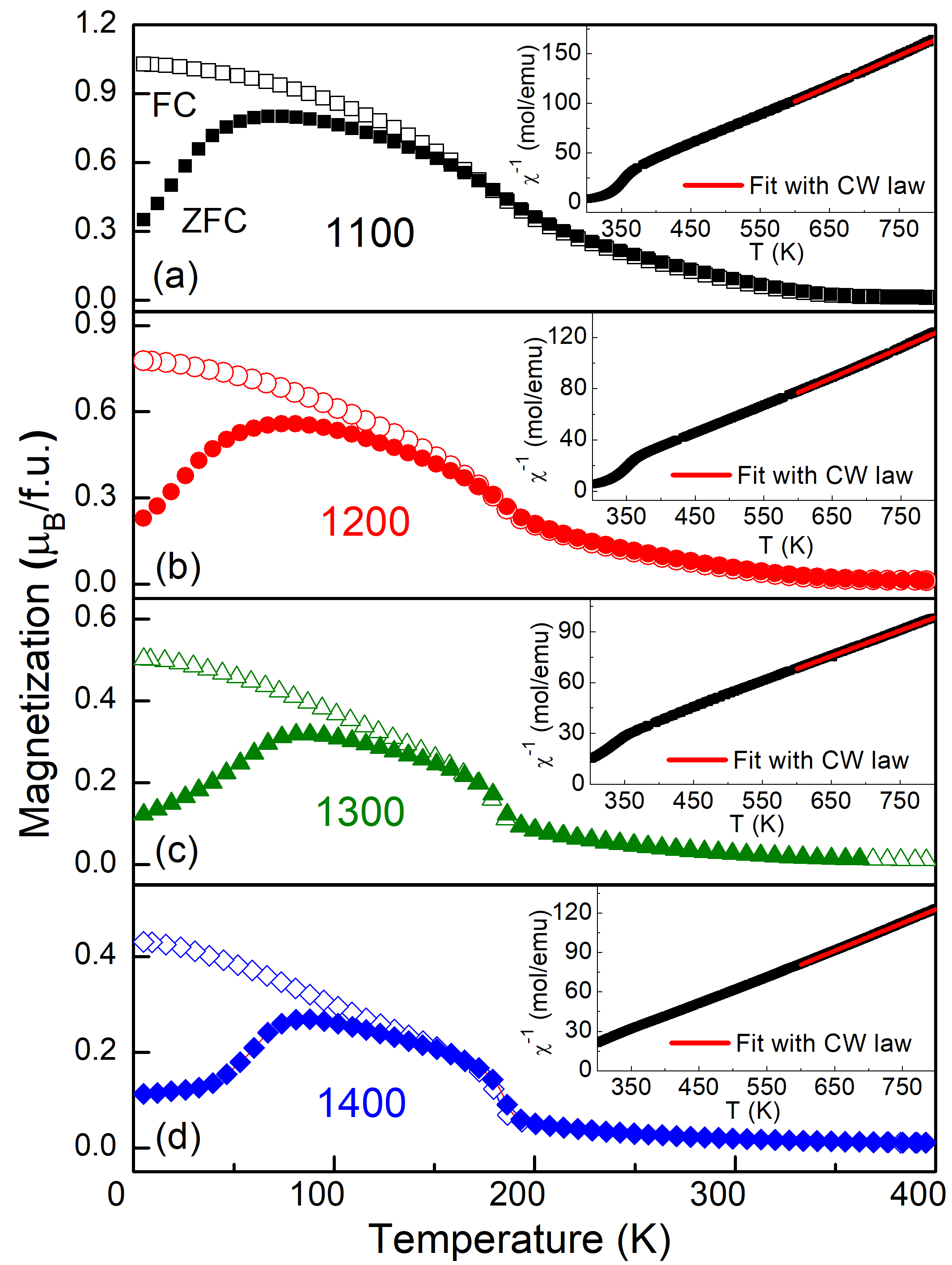}
\end{center}
\caption{ZFC-FC $M(T)$ curves measured with $H$ = 2.5 kOe for LSCMO samples (a) 1100, (b) 1200, (c) 1300, and (d) 1400. The insets show their respective $\chi^{-1}$ ($H$ = 2.5 kOe) curves measured up to 800 K, where the straight lines represent the best fit with the Curie-Weiss law.}
\label{Fig_MxT}
\end{figure}

The DC ZFC-FC magnetization against temperature [$M(T)$] curves measured with $H$ = 2.5 kOe, Fig. \ref{Fig_MxT}, show an FM-like transition below 200 K for all investigated samples, most likely associated with the Co$^{2+}$--O--Mn$^{4+}$ coupling \cite{Murthy,PRB2019,Dass}. The ZFC and FC curves bifurcate at lower temperatures, a characteristic feature of disordered polycrystalline compounds exhibiting phase competition, especially those presenting glassy magnetism \cite{Sami,Rev_SG}.

Another essential detail to be noticed in the $M(T)$ curves is that the magnetization values are much smaller than expected for an FM coupling between Co and Mn, and it systematically decreases with increasing $T_s$. For LSCMO, apart from the Co$^{2+}$--O--Mn$^{4+}$ FM coupling usually found for the pristine La$_2$CoMnO$_6$ compound, the 25\% of Sr$^{2+}$ to La$^{3+}$ substitution leads to the Co$^{2+}$/Co$^{3+}$ and Mn$^{3+}$/Mn$^{4+}$ mixed valences observed on the XANES spectra that, together with the antisite disorder, turn on other relevant magnetic interactions such as Co$^{2+/3+}$--O--Co$^{2+/3+}$, Mn$^{3+/4+}$--O--Mn$^{3+/4+}$, and Co$^{2+/3+}$--O--Mn$^{3+/4+}$ \cite{Murthy,Murthy2,PRB2019,La2-xCax,APL}. Furthermore, the presence of antiphase boundaries, in which neighboring FM domains couple antiparallel, is often invoked to explain the small magnetization of CoMn-based double-perovskite oxides \cite{single_crystal,Dass}. For several TM-based polycrystalline oxides, small grains usually represent single domains. In contrast, for larger grains, there is a tendency for each grain to be subdivided into two or more domains \cite{Zaag,Randall}. Therefore, for the case of the LSCMO samples investigated here, the decrease of magnetization with grain size enhancement may be explained by a likely concomitant increase in the number of intragrain antiphase boundaries.

The inverse of DC magnetic susceptibility ($\chi^{-1}$) curves, measured at $H=2.5$ kOe, depicted in the insets of Fig. \ref{Fig_MxT} reveal nonlinear behavior for a wide temperature interval above the FM transition temperature, suggesting that short-range correlations become relevant above the FM transition. This phenomenon, already found for similar CoMn-based double-perovskite oxides \cite{JPCM,Pal}, is observed for materials presenting the so-called Griffiths phase, for which FM clusters start to develop in the paramagnetic matrix well above $T_C$, and their volume increases continuously as temperature decreases until the FM coupling finally percolates around $T_C$ \cite{Griffiths,Bray}. It is also possible that this anomaly in the $\chi^{-1}$ curves is related to some other type of short-range correlation associated with magnetic inhomogeneity \cite{He}, or even with Co$^{3+}$ spin state transition \cite{Abbate}. A detailed study regarding the above $T_C$ properties of these LSCMO samples is in progress.

For temperatures above 600 K, the linearity of the $\chi^{-1}$ curves is undoubtedly established. From the fits of the purely paramagnetic regions with the Curie-Weiss (CW) law, we observe that the Curie-Weiss temperature, $\theta_{CW}$, is largely positive for all samples (see Table \ref{T2}), indicating that the FM coupling is dominant, in agreement with previous reports for LSCMO and resemblant CoMn-based compounds \cite{Murthy,Giri,Dass}. Table \ref{T2} also reveals a decrease in $\theta_{CW}$ with increasing $T_s$, which, at first glance, could suggest a progressive weakening in the coupling between the TM ions. However, since a systematic decrease of $T_C$ does not accompany this trend, it most likely indicates that AFM interactions become relatively more relevant as the grain size increases.

The $K$-edge XANES data indicate the same Co and Mn valence states for all the LSCMO samples here investigated (see Fig. S1 of the Supplementary Material \cite{SM}). Thus, one can expect nearly the same effective magnetic moments ($\mu_{eff}$) for these samples. Indeed, from the Curie-Weiss fits, we obtain $\mu_{eff}$ = 5.5 $\mu_B$/f.u. for all the samples except 1100, which has the lower occupation of the transition-metal ions at their site. The fact that $\mu_{eff}$ is the same gives evidence that the Co and Mn valences do not have any dependence for $T_s > 1100^{\circ}$C, in agreement with what was observed from XANES. We can estimate the valence states of the transition-metal ions present in these samples by comparing the $\mu_{eff}$ experimentally observed with the theoretical value calculated by the following equation for systems with two or more transition-metal ions
\begin{equation}
\mu = \sqrt{{\mu_1}^2 + {\mu_2}^2 + {\mu_3}^2...}. \label{EqMU}
\end{equation}
Assuming that the valence state of Co and Mn are approximately Co$^{2.4+}$ and Mn$^{3.9+}$, as previously observed for LSCMO employing $L_{2,3}$-edge X-ray absorption spectroscopy \cite{PRB2019}, and using the standard individual magnetic moments for the transition-metal ions present ($\mu_{Co^{2+}}$ $\simeq$ 4.8 $\mu_B$, $\mu_{Co^{3+}}$ $\simeq$ 5.4 $\mu_B$, $\mu_{Mn^{3+}}$ $\simeq$ 5 $\mu_B$, $\mu_{Mn^{4+}}$ $\simeq$ 4 $\mu_B$ \cite{Ashcroft}), we obtain $\mu$ = 6.5 $\mu_B$/f.u., which is somewhat larger than the experimental value. On the other hand, assuming the low-spin configuration for Co$^{3+}$ ($\mu_{LSCo^{3+}}$ = 0) yields $\mu$ = 5.5 $\mu_B$/f.u., precisely the same as the experiment. Indeed, this is just a rough estimate since other parameters, such as the off-stoichiometry of oxygen and transition-metal ions and deviations from the here-assumed individual magnetic moments, certainly affect the system's effective moment. In any case, our results suggest a low-spin configuration for Co$^{3+}$, and the presence of this nonmagnetic ion may play its part in the uncompensated exchange interactions responsible for the EB effect \cite{PRB2019,JPCM}.

\begin{table}
\caption{Main results obtained from the ZFC-FC DC $M(T)$, $\chi_{AC}$, IRM, and ZFC $M(H)$ curves.}
\label{T2}
\begin{tabular}{c|cccc}
\hline \hline
$T_s$ ($^{\circ}$C) & 1100 & 1200 & 1300 & 1400 \\
\hline
$T_C$ (K) & 187.2 & 183.2 & 184.8 & 185.1 \\
$T_2$ (K) & 152.8 & 148.0 & 145.2 & 143.5 \\
$\theta_{CW}$ (K) & 303 & 298 & 276 & 270 \\
$\mu_{eff}$ ($\mu$/f.u.) & 4.7 & 5.5 & 5.5 & 5.5 \\
\hline
$T_g$ (K) & 59.2 & 65.1 & 69.0 & 71.9 \\
$\tau_0$ (s) & 1.0$\times$10$^{-5}$ & 1.5$\times$10$^{-6}$ & 1.7$\times$10$^{-7}$ & 8.6$\times$10$^{-8}$ \\
$z\nu$ & 6.9 & 6.2 & 7.3 & 7.6 \\
$\delta T_f$ & 0.095 & 0.081 & 0.071 & 0.065 \\
\hline
$M_{sp}$ & 0.56 & 0.59 & 0.65 & 0.68 \\
$M_g$ & 0.44 & 0.41 & 0.35 & 0.32 \\
$t_p$ (s) & 1.6$\times$10$^8$ & 9.9$\times$10$^8$ & 4.2$\times$10$^{10}$ & 4.5$\times$10$^{10}$ \\
$n$ & 0.070 & 0.063 & 0.086 & 0.089 \\
\hline
$H_{EB}$ (Oe) & 279 & 783 & 3545 & 4271 \\
$H_C$ (Oe) & 2696 & 3719 & 8002 & 10245 \\
\hline \hline
\end{tabular}
\end{table}

\begin{figure}
\begin{center}
\includegraphics[width=0.5 \textwidth]{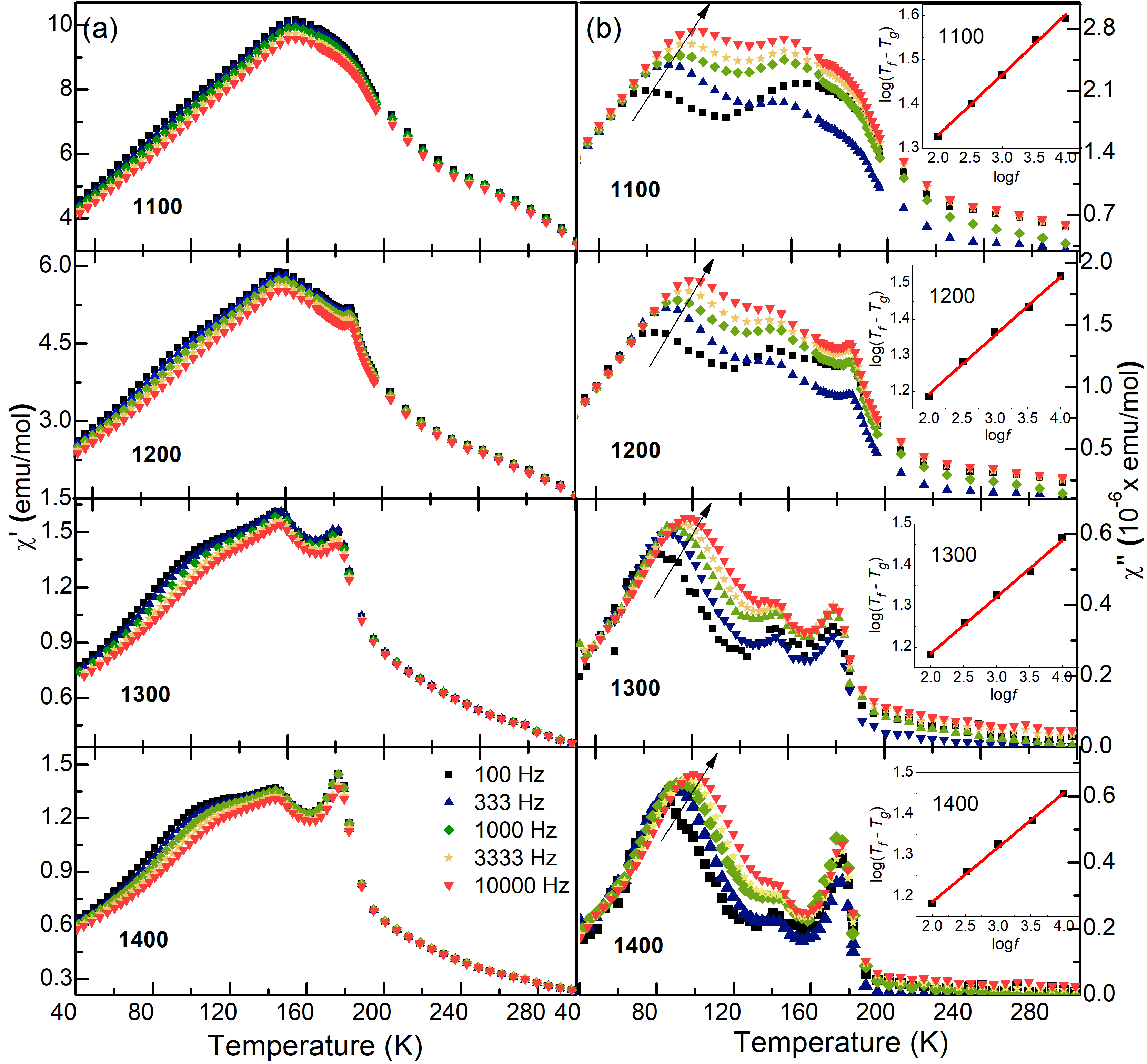}
\end{center}
\caption{(a) $\chi'$ and (b) $\chi''$ as a function of temperature for the LSCMO samples, measured with $H_{AC}$ = 5 Oe at five different frequencies. The insets show the evolution of $T_f$ with the frequency for each compound, where the solid lines represent the fits with Eq. \ref{EqCHIac}.}
\label{Fig_chiAC}
\end{figure}

To investigate the spin dynamics in the LSCMO samples and further verify how the multiple valence states affect its magnetization, we measured AC magnetic susceptibility ($\chi_{AC}$) as a function of temperature, with oscillating field $H_{AC}$ = 5 Oe and five frequencies ranging from 100 Hz to 10000 Hz. The results in Fig. \ref{Fig_chiAC} reveal some unnoticed anomalies in the DC curves. The higher temperature peak ($T_1$) seen on the real part of $\chi_{AC}$ ($\chi'$) corresponds to the Co$^{2+}$--O--Mn$^{4+}$ coupling, while the second one ($T_2$) gives further evidence of the effect of multiple valence states on LSCMO. It could be related to Co$^{2+/3+}$--O--Co$^{2+/3+}$, Mn$^{3+/4+}$--O--Mn$^{3+/4+}$ AFM interactions, Mn$^{3+}$--O--Mn$^{4+}$ FM interaction, or even vibronic Co$^{3+}$--O--Mn$^{3+}$ FM coupling, as already suggested in literature \cite{PRB2019,Dass}. It is important to notice that there are some subtle changes in the magnitude of these first two peaks, but their horizontal positions do not change with the frequency, demonstrating that they are associated with ordinary magnetic transitions \cite{Fujiki,Balanda}. On the other hand, a somewhat more prominent change in the magnetization with frequency at lower temperatures is noticed, as well as the appearance of a rounded peak that becomes more evident in the higher $T_s$ compounds. The presence of this anomaly is more apparent in the imaginary part of $\chi_{AC}$ ($\chi''$), Fig. \ref{Fig_chiAC}(b), where it shifts toward higher temperatures as the frequency increases. This is a signature of glassy magnetic behavior, as expected for a disordered system presenting competing magnetic phases \cite{Rev_SG,PRB2019,Murthy2}. For all the samples, the frequency-dependent freezing temperature ($T_f$) could be well described by the critical slowing down model of the dynamic scaling theory \cite{Hohenberg,Souletie}, which predicts the following power law relation between $T_f$ and the frequency
\begin{equation}
\frac{\tau}{\tau_{0}}=\left[\dfrac{(T_{f} - T_{g})}{T_{g}}\right]^{-z\nu}, \label{EqCHIac}
\end{equation}
where $\tau$ = 1/$f$ is the inverse of the frequency, $\tau_0$ is a parameter corresponding to the characteristic relaxation time of spin-flip, $T_{g}$ is the glassy transition temperature as the frequency tends to zero, $z$ is the dynamical critical exponent and $\nu$ is the critical exponent of the correlation length. The solid lines at the insets of Fig. \ref{Fig_chiAC}(b) represent the best fits to Eq. \ref{EqCHIac}. The $\tau_0$ and $z\nu$ values obtained, depicted in Table \ref{T2}, are typical of cluster SG systems \cite{Murthy2,Souletie,Malinowski}. Moreover, the decrease of $\tau_0$ with increasing grain size is noticed, hinting at a tendency toward canonical SG behavior as $T_s$ increases \cite{Souletie,Anand2}.

Another criterion that is often used to classify the material as canonical SG, cluster SG or superparamagnet is the so-called Mydosh's parameter, $\delta T_f$ = $\Delta T_f/T_f(\Delta log f)$ \cite{Mulder}, where for canonical SG $\delta T_f$ $\lesssim$ 0.01, for superparamagnets $\delta T_f$ $\gtrsim$ 0.1, and for cluster SG it has intermediate values between those of SG and superparamagnets \cite{Murthy2,Souletie,Malinowski,Anand2}. The $\delta T_f$ values here obtained classify the samples as cluster SG (Table \ref{T2}), and its systematic decrease confirms the tendency toward canonical SG with increasing the $T_s$.

Since previous studies propose that the SEB of LSCMO and similar double-perovskite oxides are related to the dynamics of relaxation of the SG-like moments under the effect of $H$ \cite{Model,Model2}, we have measured isothermal remanent magnetization (IRM) curves for the samples of interest. The protocol to obtain the curves was as follows: each sample was ZFC down to 5 K, then $H$ was continuously increased up to 90 kOe at a constant $H$ sweep rate of 100 Oe/s, followed by its continuous decrease down to zero at the same sweep rate. Immediately after $H$ reached zero, the material's remanent magnetization was captured as a function of time ($t$). Importantly, this protocol of increasing $H$ up to 90 kOe followed by its decrease to zero corresponds to the measurement of the first quadrant of a $M(H)$ curve.

\begin{figure}
\begin{center}
\includegraphics[width=0.48 \textwidth]{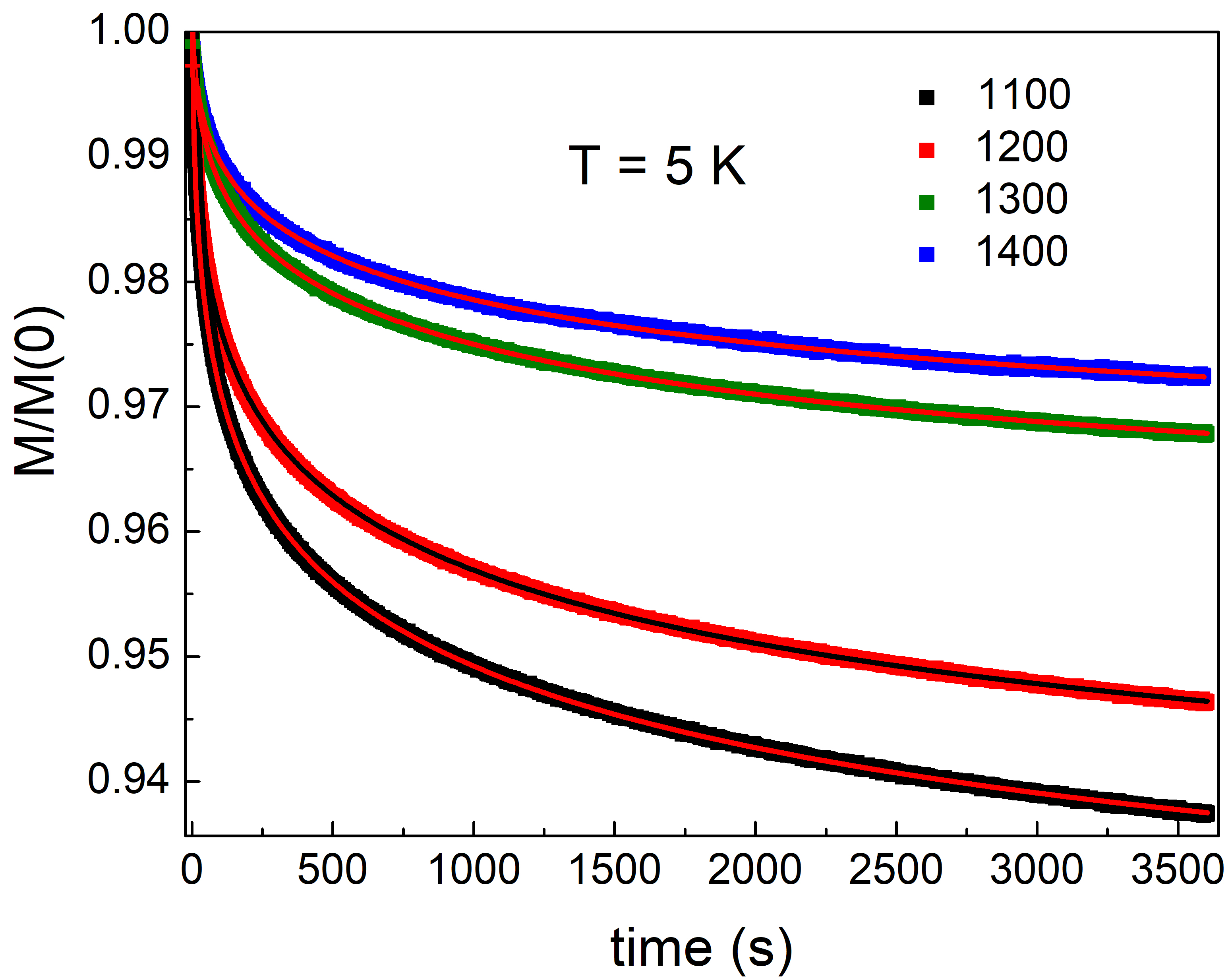}
\end{center}
\caption{IRM curves measured at 5 K on the LSCMO samples, normalized by their $t$ = 0 values. The solid lines represent the fits with Eq. \ref{EqIRM}.}
\label{Fig_MxTime}
\end{figure}

The IRM curves of an SG-like system are known by their long-lasting decay, where the remanent magnetization usually obeys a stretched exponential equation \cite{Rev_SG,Nordblad}. In our case, LSCMO is a reentrant spin glass system presenting its SG-like phase concomitant to other conventional magnetic phases. Therefore, the IRM curves can be fitted by the following equation,
\begin{equation}
M(t)=M_{sp} + M_{g}e^{-(t/t_{p})^{n}}, \label{EqIRM}
\end{equation}
where $M_{sp}$ corresponds to the spontaneous magnetization of the ordinary FM phase, $M_g$ is the initial magnetization of the SG-like phase, $t_{p}$ and $n$ are the time and the time-stretch exponential constants, respectively. Fig. \ref{Fig_MxTime} shows the IRM curves normalized by the magnetization value at $t$ = 0. From the data in Fig. \ref{Fig_MxTime}, we can note that the magnetization's decay gets slower as $T_s$ increases. Besides, from the results of the fits shown in Table \ref{T2}, we observe that the SG-like phase's relative fraction decreases with increasing $T_s$. These results will have an impact on the SEB effect, as will be discussed next.

\begin{figure*}
\begin{center}
\includegraphics[width= \textwidth]{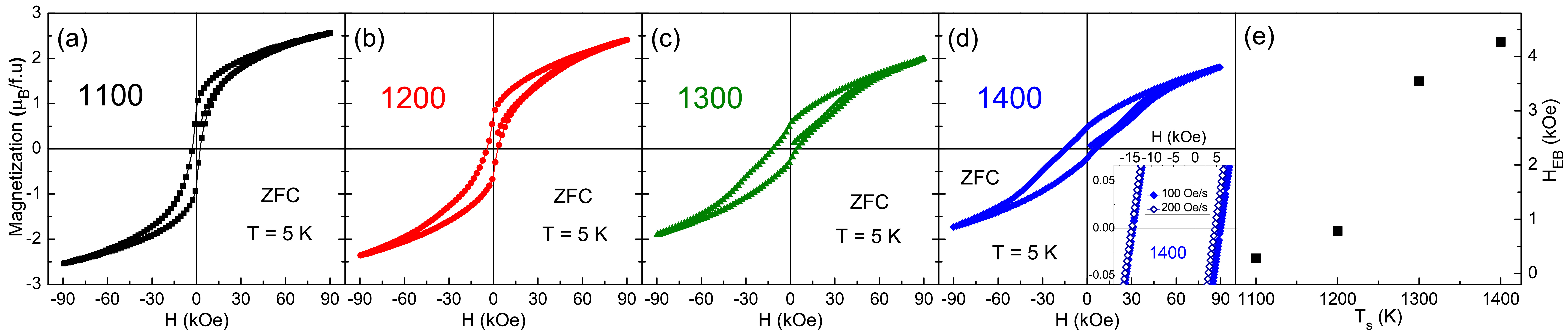}
\end{center}
\caption{ZFC $M(H)$ loops measured at 5 K on samples (a) 1100, (b) 1200, (c) 1300 and (d) 1400. The inset in panel (d) shows magnified views of the coercive fields for $M(H)$ curves measured on sample 1400 with field sweep rates of 100 and 200 Oe/s. (e) Evolution of $H_{EB}$ with $T_s$.}
\label{Fig_MxH}
\end{figure*}

As aforementioned, the EB effect is manifested as a shift in the $M(H)$ curves. The EB field, defined as $H_{EB}$ = $|H^{+} + H^{-}|$/2 where $H^{+}$ and $H^{-}$ are, respectively, the coercive fields at the ascending and descending branches of the hysteresis loop, gives a measure of the magnitude of the UA. The average coercive field is calculated as $H_C$ = ($H^{+}$ - $H^{-}$)/2. Fig. \ref{Fig_MxH} displays the $M(H)$ curves measured at an $H$ sweep rate of 100 Oe/s, after ZFC the samples down to 5 K. As Fig. \ref{Fig_MxH}(e) and Table \ref{T2} show, $H_{EB}$ and $H_C$ systematically increase with $T_s$. The inset of Fig. \ref{Fig_MxH}(d) gives a comparative view of the asymmetry of the curves measured for sample 1400 at two distinct $H$ sweep rates, 100 and 200 Oe/s. As can be noticed, the SEB effect depends on the sweep rate, further evidencing that the relaxation of the SG-like moments plays a role in it \cite{Model,Model2}.

\section{Discussion}

Our IRM curves show a decrease in the fraction of SG-like phase present in the LSCMO samples as $T_s$ increases, while the SEM images reveal the increase of grain size with $T_s$. Since the surface-to-core ratio naturally decreases as the grain size increases, these results suggest that the SG-like phase might be in the surfaces of the LSCMO grains, as often found on polycrystalline EB systems \cite{Nogues,Nogues2,Maity,Kodama,Bedanta}. Our previous study corroborates this scenario, demonstrating that a single-crystalline sample of LSCMO does not show SG-like (nor SEB) behavior \cite{single_crystal}.

From the above and given that the SG-like phase seems to be a ubiquitous ingredient of SEB on double perovskites \cite{Maity,Murthy,Murthy2,Pal,Giri}, at first glance, one could guess that $H_{EB}$ should decrease with the increase of the grain size, \textit{i.e.} with the decrease in the fraction of the glassy magnetic phase. However, the opposite trend is observed here, ruling out this possibility.

Here, we recall that the EB phenomenon is known as an interface effect, with the exchange coupling at the interface between distinct magnetic phases giving rise to pinned moments responsible for the UA \cite{Nogues,Nogues2}. In this sense, the SEM images depicted in Figs. \ref{Fig_SEM} (and Fig. S12 of the Supplementary Material \ref{SM}) show small and isolated grains in the samples produced at lower $T_s$. Indeed, there is no interface coupling between the isolated grains since superexchange is an atomistic exchange interaction that persists only up to a few angstroms.  Although impingement between some grains can be observed, for which the defective and frustrated character of the coalesced regions could be understood as barriers to the percolation of long-range magnetic order along adjacent grains, the application of a magnetic field during the $M(H)$ cycle, however, may favor the alignment of the initially frustrated moments toward field direction, signifying a field-induced collective behavior where the two impinged grains may be now roughly viewed as a single domain.

As Fig. \ref{Fig_SEM} also demonstrates, besides the increase of average grain size that occurs with enhancing $T_s$, there is a simultaneous tendency of formation of faceted grains, where a clear distinction between adjacent grains is established. The intergrain region that connects adjacent facets may still be defective and frustrated, \textit{i.e.} it is an SG-like phase. At the same time, the systematic decrease in the magnitude of magnetization with increasing $T_s$ suggests that these intergrain regions impinged faceted interfaces may act as antiphase boundaries. Within this scenario of antiphase boundaries, typical of CoMn-based perovskites \cite{Dass}, the antiparallel alignment of adjacent grains results in competing magnetic phases at the interfaces. At the same time, the glassy region connecting the facets acts as the pinning centers necessary for the onset of the EB effect.

Our current findings suggest that $H_{EB}$ initially enhances with $T_s$ due to the strengthened exchange coupling at the intergrain interfaces. Still, at some point, the systematic decrease of the SG-like phase with $T_s$ should dominate, leading to the decrease of the SEB effect until it vanishes for the single crystal, since it was demonstrated the absence of SEB (as well as the SG-like phase) on single crystalline LSCMO \cite{single_crystal}. Here, it is worth mentioning that it was already reported SEB effect on a single-crystalline perovskite, namely the SmFeO$_3$ compound, for which magnetic force microscopy revealed the formation of cluster glasses and magnetic phase separation \cite{SmFeO3}. Although its SEB was attributed to the formation of pinning points at small parts of the clusters, it is unclear whether these parts correspond to the domain boundaries. Further investigation of this material and related compounds is necessary to verify whether the underlying physics of its exchange anisotropy is similar or fundamentally different from that of our system of interest.

At this stage, it is not possible to precisely determine the magnetic texture at the interfaces of adjacent grains in our LSCMO samples. Additional experimental techniques, such as X-ray tomography, are necessary to unravel this issue. Nevertheless, regarding the polycrystalline SEB materials, our results attest to the fact that an SG-like phase is an essential ingredient of SEB. Additionally, it shows that this is not the only condition for the emergence of spontaneous UA. There must be some minimal degree of coupling between the grain's interfaces for the SEB to become noticeable. This makes designing strategies to tune the SEB effect with adequate sintering conditions possible. Also important, it helps to explain why, albeit all SEB compounds are reentrant spin glasses, not all reentrant spin glass materials exhibit SEB.

\section{Summary}
Here, we produced four polycrystalline LSCMO samples presenting different grain sizes by sintering them at distinct $T_s$. While the PXRD and PND data show that there are no significant changes in the lattice parameters nor the Co/Mn-site occupation, SEM images demonstrate the increase of grain size with $T_s$, with the samples presenting larger grain size exhibiting faceted morphology. The Co and Mn $K$-edge XANES spectra indicate the same formal valence states for the transition metals on all samples. In contrast, the DC $M(T)$ curves indicate the presence of multiple magnetic phases that, together with the antisite disorder at the Co/Mn site, lead to reentrant spin glass behavior at low temperatures. The $\chi_{AC}$ and IRM curves indicate that the SG-like phase lies in the grain boundaries and, although the fraction of the glassy magnetic phase decreases with increasing $T_s$, the antiphase formed between adjacent grains establishes the presence of competing magnetic phases at the faceted interfaces, necessary to the onset of EB. Our experimental results demonstrate an effective method for synthetically tuning the SEB effect in the LSCMO family of materials, and this synthetic control over SEB may be generalizable.

\begin{acknowledgements}
This work was supported by the Brazilian funding agencies: Funda\c{c}\~{a}o Carlos Chagas Filho de Amparo \`{a} Pesquisa do Estado do Rio de Janeiro (FAPERJ) [Nos. E-26/202.798/2019 and E-26/211.291/2021], Funda\c{c}\~{a}o de Amparo \`{a}  Pesquisa do Estado de Goi\'{a}s (FAPEG) and Conselho Nacional de Desenvlovimento Cient\'{\i}fico e Tecnol\'{o}gico (CNPq). This research used facilities of the Brazilian Synchrotron Light Laboratory (LNLS), part of the Brazilian Center for Research in Energy and Materials (CNPEM), a private non-profit organization under the supervision of the Brazilian Ministry for Science, Technology, and Innovations (MCTI). The EMA beamline staff is acknowledged for their assistance during the experiments 20220583. Certain commercial equipment, instruments, or materials are identified in this document. Such identification does not imply recommendation or endorsement by the National Institute of Standards and Technology, nor does it imply that the products identified are necessarily the best available for the purpose. The views expressed in the article do not necessarily represent the views of the DOE or the U.S. Government. The U.S. Government retains, and the publisher, by accepting the article for publication, acknowledges that the U.S. Government retains a nonexclusive, paid-up, irrevocable, worldwide license to publish or reproduce the published form of this work or allow others to do so for U.S. Government purposes. A portion of this research used resources at the Spallation Neutron Source, a DOE Office of Science User Facility operated by the Oak Ridge National Laboratory. A portion of this work was supported by NIST. R.A.K. gratefully acknowledges from the U.S. DOE Office of Energy Efficiency and Renewable Energy (EERE), Hydrogen and Fuel Cell Technologies Office (HFTO) contract no. DE-AC36-8GO28308 to the National Renewable Energy Laboratory (NREL).
\end{acknowledgements}


\begin{thebibliography}{99}

\bibitem{Rao} C. N. R. Rao. Transition metal oxides, Annu. Rev. Phys. Chem. \textbf{40}, 291-326 (1989).

\bibitem{Tokura} Y. Tokura and N. Nagaosa. Orbital Physics in Transition-Metal Oxides, Science \textbf{288}, 462 (2000).

\bibitem{Sami} S. Vasala and M. Karppinen. $A_{2}B'B''O_{6}$ perovskites: A review,  Prog. Solid State Chem. \textbf{43}, 1 (2015).

\bibitem{Nogues} J. Nogu\'{e}s and I. K. Schuller. Exchange bias, J. Magn. Magn. Mater. \textbf{192}, 203 (1999).

\bibitem{Meiklejohn} W. H. Meiklejohn and C. P. Bean. New Magnetic Anisotropy, Phys. Rev. \textbf{102}, 1413 (1956).

\bibitem{Nogues2} J. Nogu\'{e}s, J. Sort, V. Langlais, V. Skumryev, S. Suri\~{n}ach, J. S. Mu\~{n}oz, M. D. Bar\'{o}. Exchange bias in nanostructures, Phys. Rep. \textbf{422}, 65-117 (2005).

\bibitem{Wang} B. M. Wang, Y. Liu, P. Ren, B. Xia, K. B. Ruan, J. B. Yi, J. Ding, X. G. Li, and L. Wang. Large Exchange Bias after Zero-Field Cooling from an Unmagnetized State, Phys. Rev. Lett. \textbf{106}, 077203 (2011).

\bibitem{Maity} T. Maity, S. Goswami, D. Bhattacharya, and S. Roy. Superspin Glass Mediated Giant Spontaneous Exchange Bias in a Nanocomposite
of BiFeO$_3$-Bi$_2$Fe$_4$O$_9$, Phys. Rev. Lett. \textbf{110}, 107201 (2013).

\bibitem{Model} L. T. Coutrim, E. M. Bittar, F. Garcia, and L. Bufai\c{c}al. Influence of spin glass-like magnetic relaxation on the zero-field-cooled exchange bias effect, Phys. Rev. B \textbf{98}, 064426 (2018).

\bibitem{Model2} L. Bufai\c{c}al, L. T. Coutrim, E. M. Bittar, F. Garcia. A phenomenological model for the spontaneous exchange bias effect, J. Magn. Magn. Mater. \textbf{512}, 167048 (2020).

\bibitem{Rev_SG} K. Binder and A. P. Young. Spin glasses: Experimental facts, theoretical concepts, and open questions, Rev. Mod. Phys. \textbf{58}, 801 (1986).

\bibitem{Murthy} J. Krishna Murthy and A. Venimadhav. Giant zero field cooled spontaneous exchange bias effect in phase-separated La$_{1.5}$Sr$_{0.5}$CoMnO$_6$, Appl. Phys. Lett. \textbf{103}, 25410 (2013).

\bibitem{Murthy2} J. Krishna Murthy, K. D. Chandrasekhar, H. C. Wu, H. D. Yang, J. Y. Lin and A. Venimadhav. Antisite disorder driven spontaneous exchange bias effect in La$_{2-x}$Sr$_x$CoMnO$_6$ (0 $\leq x\leq$ 1), J. Phys.: Condens. Matter 28, 086003 (2016).

\bibitem{La2-xCax} J. R. Jesus, L. Bufai\c{c}al, E. M. Bittar. The spontaneous exchange bias effect in La$_{2-x}$Ca$_x$CoMnO$_6$ series, J. Magn. Magn. Mater. 556, 169402 (2022).

\bibitem{PRB2019} L. T. Coutrim, D. Rigitano, C. Macchiutti, T. J. A. Mori, R. Lora-Serrano, E. Granado, E. Sadrollahi, F. J. Litterst, M. B. Fontes, E. Baggio-Saitovitch, E. M. Bittar, and L. Bufai\c{c}al. Zero-field-cooled exchange bias effect in phase-segregated La$_{2-x}A_x$CoMnO$_{6-\delta}$ ($A$ = Ba,Ca,Sr; $x$ = 0, 0.5), Phys. Rev. B 100, 054428 (2019).

\bibitem{APL} M. Boldrin, A. G. Silva, L. T. Coutrim, J. R. Jesus, C. Macchiutti, E. M. Bittar, and L. Bufai\c{c}al. Tuning the spontaneous exchange bias effect with Ba to Sr partial substitution in La$_{1.5}$(Sr$_{0.5-}$Ba$_x$)CoMnO$_6$, Appl. Phys. Lett. 117, 212402 (2020).

\bibitem{La2-xBax} H. Fabrelli, A. G. Silva, M. Boldrin, L. Bufai\c{c}al, E. M. Bittar. Structural transitions and spontaneous exchange bias in La$_{2-x}$Ba$_x$CoMnO$_6$ series, J. Solid State Chem. \textbf{322}, 123944 (2023).

\bibitem{Zhang} H. G. Zhang, L. Xie, X. C. Liu, M. X. Xiong, L. L. Cao and Y. T. Li. The reversal of the spontaneous exchange bias effect and zero-field-cooling magnetization in La$_{1.5}$Sr$_{0.5}$Co$_{1-x}$Fe$_x$MnO$_6$: the effect of Fe doping, Phys. Chem. Chem. Phys. \textbf{19}, 25186 (2017).

\bibitem{JPCM} A. G. Silva, K. L. Salcedo Rodr\'{i}guez, C. P. Contreras Medrano, G. S. G. LourenÃ§o, M. Boldrin, E Baggio-Saitovitch and L Bufai\c{c}al. Griffiths phase and spontaneous exchange bias in La$_{1.5}$Sr$_{0.5}$CoMn$_{0.5}$Fe$_{0.5}$O$_6$, J. Phys.: Condens. Matter \textbf{33}, 065804 (2021).

\bibitem{Zhao} Hongguang Zhang, Wei Chen, Liang Xie, HuiHui Zhao, Qi Li. Tunable exchange bias in La$_{1.5}$Sr$_{0.5}$CoMnO$_6$ double perovskite doped with nonmagnetic Ga ions, Curr. Appl. Phys. \textbf{35}, 58-66 (2022).

\bibitem{Pal} Arkadeb Pal, Prajyoti Singh, V. K. Gangwar, Amish G. Joshi, P. Khuntia, G. D. Dwivedi, Prince K. Gupta, Mohd Alam, Khyati Anand, K. Sethupathi, Anup K. Ghosh and Sandip Chatterjee. Probing the Griffiths-like phase, unconventional dual glassy states, giant exchange bias effects, and its correlation with its electronic structure in Pr$_{2-x}$Sr$_x$CoMnO$_6$, J. Phys.: Condens. Matter \textbf{32}, 215801 (2020).

\bibitem{Giri} S. K. Giri, R. C. Sahoo, Papri Dasgupta, A. Poddar and T. K. Nath. Giant spontaneous exchange bias effect in Sm$_{1.5}$Ca$_{0.5}$CoMnO$_6$ perovskite, J. Phys. D: Appl. Phys. \textbf{49}, 165002 (2016).

\bibitem{single_crystal} C. Macchiutti, J. R. Jesus, F. B. Carneiro, L. Bufai\c{c}al, M. Ciomaga Hatnean, G. Balakrishnan, and E. M. Bittar. Absence of zero-field-cooled exchange bias effect in single crystalline La$_{2-x}A_x$CoMnO$_6$($A$ = Ca, Sr) compounds, Phys. Rev. Mater. \textbf{5}, 094402 (2021).

\bibitem{TOPAS} A. Coelho, Topas Academic v6; Coelho Softw. (2017). http://www.topas-academic.net/.

\bibitem{CMPR} B. H. Toby, CMPR - a powder diffraction toolkit. J. Appl. Cryst. \textbf{38}, 1040 (2005).

\bibitem{LNLS} R. R. Geraldes, S. A. L. Luiz, J. L. de Brito Neto, T. R. S. Soares, R. D. dos Reis, G. A. Calligaris, G. Witvoetb, and J. P. M. B. Vermeulen. Fly-scan-oriented motion analyses and upgraded beamline integration architecture for the High-Dynamic Double-Crystal Monochromator at Sirius/LNLS, J. Synchrotron Rad. \textbf{30}, 90 (2023).

\bibitem{SM} See Supplemental Material at \href{http://link.aps.org/supplemental/10.1103/PhysRevMaterials.8.044408}{PhysRevMaterials.8.044408} for details of XANES experiments and data and the PXRD and PND Rietveld refinements for all investigated samples.

\bibitem{Serrate} D. Serrate, J. M. De Teresa, and M. R. Ibarra. Double perovskites with ferromagnetism above room temperature, J. Phys.: Condens. Matter \textbf{19}, 023201 (2007).

\bibitem{Sears} V. F.Sears. Neutron Scattering Lengths and Cross Sections, Neutron News \textbf{3}, 26 (1992).

\bibitem{Bera} Souhardya Bera, Ankit Saha, Shibsankar Mondal, Arnab Biswas, Shreyasi Mallick, Rupam Chatterjee and Subhasis Roy. Review of defect engineering in perovskites for photovoltaic application, Mater. Adv. \textbf{3}, 5234 (2022).

\bibitem{Dass} R. I. Dass and J. B. Goodenough. Multiple magnetic phases of La$_2$CoMnO$_{6-\delta}$ (0 $\leq \delta \leq$ 0.05), Phys. Rev. B \textbf{67}, 014401 (2003).

\bibitem{Zaag} P. J. van der Zaag, P. J. van der Valk, M. Th. Rekveldt. A domain size effect in the magnetic hysteresis of NiZn-ferrites,  Appl. Phys. Lett. \textbf{69}, 2927-2929 (1996).

\bibitem{Randall} Wenwu Cao, Clive A. Randall. Grain size and domain size relations in bulk ceramic ferroelectric materials, J. Phys. Chem. Solids \textbf{57}, 10, 1499-1505 (1996).

\bibitem{Griffiths} R. B. Griffiths. Nonanalytic behavior above the critical point in a random Ising ferromagnet, Phys. Rev. Lett. \textbf{23}, 17 (1969).

\bibitem{Bray} A. J. Bray. Nature of the Griffiths Phase, Phys. Rev. Lett. \textbf{59}, 586 (1987).

\bibitem{He} C. He, M. A. Torija, J. Wu, J. W. Lynn, H. Zheng, J. F. Mitchell, and C. Leighton. Non-Griffiths-like clustered phase above the Curie temperature of the doped perovskite cobaltite La$_{1-x}$Sr$_x$CoO$_3$, Phys. Rev. B \textbf{76}, 014401 (2007).

\bibitem{Abbate} M. Abbate, J. C. Fuggle, A. Fujimori, L. H. Tjeng, C. T. Chen, R. Potze, G. A. Sawatzky, H. Eisaki, and S. Uchida. Electronic structure and spin-state transition of
LaCoO$_3$, Phys. Rev. B \textbf{47}, 16124 (1993).

\bibitem{Ashcroft} N. W. Ashcroft and N. D. Mermin. Solid State Physics, Brooks/Cole, New York (1976
).

\bibitem{Fujiki} S. Fujiki and S. Katsura. Nonlinear Susceptibility in the Spin Glass, Prog. Theor. Phys. \textbf{65}, 4 (1981).

\bibitem{Balanda} M. Balanda. AC Susceptibility Studies of Phase Transitions and Magnetic Relaxation: Conventional, Molecular and Low-Dimensional Magnets, Acta Phys. Pol. A \textbf{124}, 6 (2013).

\bibitem{Hohenberg} P. C. Hohenberg and B. I. Halperin. Theory of dynamic critical phenomena, Rev. Mod. Phys. \textbf{49}, 3 (1977).

\bibitem{Souletie} J. Souletie and J.L. Tholence. Critical slowing down in spin glasses and other glasses: Fulcher versus power law, Phys. Rev. B \textbf{32}, 1 (1985).

\bibitem{Malinowski} A. Malinowski, V. L. Bezusyy, R. Minikayev, P. Dziawa, Y. Syryanyy, and M. Sawicki. Spin-glass behavior in Ni-doped La$_{1.85}$Sr$_{0.15}$CuO$_4$, Phys. Rev. B \textbf{84}, 024409 (2011).

\bibitem{Anand2} V. K. Anand, D. T. Adroja, and A. D. Hillier. Ferromagnetic cluster spin-glass behavior in PrRhSn$_3$, Phys. Rev. B \textbf{85}, 014418 (2012).

\bibitem{Mulder} C. A. M. Mulder, A. J. van Duyneveldt, J. A. Mydosh. Susceptibility of the $Cu$Mn spin-glass: Frequency and field dependences, Phys. Rev. B \textbf{23}, 1384 (1981).

\bibitem{Nordblad} P. Nordblad, P. Svedlindh, L. Lundgren, and L. Sandlund. Time decay of the remanent magnetization in a $Cu$Mn spin glass, Phys. Rev. B \textbf{33}, 645 (1986).

\bibitem{Kodama} R. H. Kodama and A. E. Berkowitz. Atomic-scale magnetic modeling of oxide nanoparticles, Phys. Rev. B \textbf{59}, 6321 (1999).

\bibitem{Bedanta} Sagarika Nayak, Palash Kumar Manna, Thiruvengadam Vijayabaskaran, Braj
Bhusan Singh, J. Arout Chelvane, Subhankar Bedanta. Exchange bias in Fe/$Ir_{20}Mn_{80}$ bilayers: Role of spin-glass like interface and bulk antiferromagnet spins, J. Magn. Magn. Mater. \textbf{499}, 166267 (2020).

\bibitem{SmFeO3} Xiao-xiong Wang, Shang Gao, Xu Yan, Qiang Li, Jun-cheng Zhang, Yun-ze Long, Ke-qing Ruand and Xiao-guang Li. Giant spontaneous exchange bias obtained by tuning magnetic compensation in samarium ferrite single crystals, Phys. Chem. Chem. Phys. \textbf{20}, 3687 (2018).

\end{thebibliography}
\end{document}